\documentclass{iau}
\usepackage{natbib}
\usepackage{multirow}
\usepackage{graphicx}

\title[VLBI astrometry of two millisecond pulsars] 
{\Large{VLBI astrometry of two millisecond pulsars}} 

\author[Z. Yan et al. ]  
{Zhen~Yan$^{1, 2, 4}$,
 Zhi-qiang~Shen$^{1,4}$,
 Jian-ping~Yuan$^{3,4}$,
 Na~Wang$^{3,4}$,
 Helge~Rottmann$^5$
 \and Walter~Alef~$ ^5$}

\affiliation{$^1$Shanghai Astronomical Observatory, Chinese Academy of
  Sciences, Shanghai 200030, China\\  email: {\tt yanzhen@shao.ac.cn}\\[\affilskip]
$^2$University of Chinese Academy of Sciences, Beijing 100039, China \\[\affilskip]
$^3$Xinjiang Astronomical Observatory, Chinese Academy of Sciences, Urumqi 830011, China\\[\affilskip]
$^4$ Key Laboratory of Radio Astronomy, Chinese Academy of Sciences, China\\[\affilskip]
$^5$Max Planck Institute for Radio Astronomy, Bonn 53121, Germany}

\pubyear{2012}
\volume{291}  
\jname{\mbox{Neutron Stars and Pulsars: Challenges and Opportunities after 80 years}}
\editors{J. van Leeuwen, ed.}
\begin{document}

\maketitle

\begin{abstract}
We present astrometric results on two millisecond pulsars, PSR~B1257+12 and
PSR~J1022+1001, as carried out through VLBI. For PSR~B1257+12, a model-independent distance
of $710_{\rm -38}^{\rm +43}$~pc and proper motion of ($\mu_{\rm
  \alpha}=46.44\pm0.08$~mas/yr, $\mu_{\rm
  \delta}=-84.87\pm0.32$~mas/yr) were obtained
from 5~epochs of VLBA and 4~epochs of EVN observations, spanning about
2 years. The two dimensional proper motion of PSR~J1022+1001
($\mu_{\rm \alpha}$$\sim$$-10.13$~mas/yr, $\mu_{\delta}$$\sim$$16.89$~mas/yr)
was also estimated, using 3~epochs of EVN observations. Based on our
results, the X-ray efficiency of PSR~B1257+12 should be in the same
range as other millisecond pulsars, and not as low as previously thought.

\end{abstract}


\firstsection 
\section{Introduction}
The distance and proper motion are fundamental and important pulsar parameters.
A model-independent distance and  proper motion measurement is especially important for millisecond pulsars (MSPs).
Firstly, MSPs are old enough to leave the Galactic disk. Model-independent pulsar distance measurements
indicate that the TC93 \citep{tac93} or  NE2001 \citep{cor02} Galactic electron density distribution model
underestimates the distances for high-latitude pulsars \citep{cbv09}. Secondly, the distance and proper
motion of a pulsar are also important parameters in the pulsar timing
observation. In the Shklovskii
effect, for example, a transverse component of this pulsar velocity gives rise to an appreciable
increase in the apparent period even if the pulsar is not slowing down \citep{shk70}. For MSPs,
$\dot{P}_{\rm Shk}$ is $\sim 10^{-19}$~s/s, comparable to their observed first order period
derivative. Furthermore, MSPs have more parameters to fit in timing observations, as most
of them have companions. If the distance and proper motion have been obtained
independently, it will be helpful for the other parameters fitting.

High precision VLBI astrometry offers a powerful way to directly measure the parallaxes and proper
motions of pulsars. With the steady progress of  VLBI observation, correlation and data processing techniques, VLBI
astrometry of some pulsars has been accomplished successfully \citep{cbs96,fgb99,bbg02,cbv09,dtb09}.

Here, we report the progress of our astrometry project on two MSPs, PSR~B1257+12
and PSR~J1022+1001, with the VLBA and EVN. PSR~B1257+12 is the first extra-solar
planetary system discovered. It has been confirmed that PSR~B1257+12 has three planets in approximately
co-planar orbits \citep{wol00}. PSR~J1022+1001 is an intermediate mass binary pulsar accompanied by
a 0.9~$M_{\odot}$ white dwarf. It lies near the ecliptic plane, so that only the component of proper
motion along the ecliptic longitude can be accurately measured  with pulsar timing method \citep{kmc99}.
For these two pulsars, the astrometry results obtained by various
methods are by now very
different (see Table~\ref{tabpara}). So, it is meaningful to perform VLBI astrometry on
these pulsars and further study their related astrophysics.
{
\renewcommand{\arraystretch}{1.2}
\begin{table}
  \centering
  \caption{The distance and proper motion of PSR~B1257+12 and PSR~J1022+1001}
  \begin{tabular}{c l l c c}
  \hline
Pulsar & Distance (pc)& Proper motion (mas/yr)& Method& Reference\\
\hline
\multirow{9}*{B1257+12}& $\sim$ 620&$-$&TC93& \cite{tac93}\\
\cline{2-5}
& \multirow{2}*{$800\pm200$}& $\mu_{\rm \alpha}=46.4\pm0.1$&
\multirow{2}*{Timing}&\multirow{2}*{\cite{wol00}}\\
& & $\mu_{\rm \delta}=-82.2\pm0.2$& & \\
\cline{2-5}
& $\sim$ 450&$-$&NE2001&  \cite{cor02}\\
\cline{2-5}
& \multirow{2}*{$-$}& $\mu_{\rm \alpha}=45.5\pm0.4$& \multirow{2}*{Timing}&\multirow{2}*{\cite{kow03}}\\
& & $\mu_{\rm \delta}=-84.7\pm0.7$& & \\
\cline{2-5}
& $660_{-130}^{+210}$&$-$&Timing& \cite{wol08}\\
\cline{2-5}
& \multirow{2}*{\textbf{$710_{\rm -38}^{\rm +43}$}}& $\mu_{\rm \alpha}=46.44\pm0.08$& \multirow{2}*{\textbf{VLBI}}&\multirow{2}*{\textbf{This work}}\\
& & $\mu_{\rm \delta}=-84.87\pm0.32$& & \\
\hline
\multirow{6}*{J1022+1001}& $\sim$ 600&$-$&TC93& \cite{tac93}\\
\cline{2-5}
&$-$& $\mu_{\rm \lambda}=-17\pm2$& Timing& \cite{kmc99}\\
\cline{2-5}
& $\sim$ 440&$-$&NE2001&  \cite{cor02}\\
\cline{2-5}
& $300_{-60}^{+100}$&$-$& Timing &\cite{hbo04}\\
\cline{2-5}
& \multirow{2}*{$-$}& $\mu_{\rm \alpha}\sim-10.13$& \multirow{2}*{\textbf{VLBI}}&\multirow{2}*{\textbf{This work}}\\
& & $\mu_{\rm \delta}\sim 16.89$& & \\
\hline
  \end{tabular}
  \label{tabpara}
\end{table}
}

\section{Observations and data reduction}
The flux density of PSR~B1257+12 and PSR~J1022+1001 is about 2 and 3~mJy at 1400~MHz, respectively.
The corresponding observing wavelength of VLBA and EVN is 21~cm and 18~cm, respectively.
Including 5 epochs of
VLBA observation and 4 epochs of EVN observation, there are 9 epochs of VLBI  observations of
PSR~B1257+12 spanning 2 years. In the VLBA observations of PSR~B1257+12, two calibrators,
J1300+1206 and J1300+141A, located on opposite sides of PSR~B1257+12 in RA direction with the separation
of $0.58^{\circ}$ and $1.61^{\circ}$, were selected as phase reference sources. In the EVN
observations of PSR~B1257+12, only J1300+141A was selected as the phase reference source.
One phase reference source at $2.96^{\circ}$ away was chosen in PSR~J1022+1001 observations
with the EVN. Only 3 of 5 epochs EVN observations of PSR~J1022+1001 were successful. The VLBA
and EVN data were correlated with NRAO-DiFX and Bonn-DiFX software correlators under the
pulsar binning mode, respectively. The data was processed with AIPS following the normal
data reduction steps of phase reference observations.

\section{Results and Discussion}
Firstly, the astrometric parameters of PSR~B1257+12 are fitted with the standard weighted least squares method with
5 degrees of freedom that astrometry measurements usually use. But, there are some systematic offset in the DEC
direction between VLBA results and EVN results. To overcome this, one more parameter
$\Delta \delta_{\rm (EVN-VLBA)}$  is added to the new data fitting. The reduced $\chi^2$ of the new fitting
is 0.67 with a fitted systematic offset $\Delta \delta_{\rm (EVN-VLBA)}$ of 1.22~mas.
The parallax fitted is $1.41\pm0.08$~mas, which corresponds to a distance $710_{\rm -38}^{\rm +43}$~pc.
The corresponding proper motion in RA and DEC direction is $46.44\pm0.08$ and $-84.87\pm0.32$ mas/yr.
The covariance between parallax ($\pi$) and proper motion ($\mu_{\rm \alpha}$, $\mu_{\rm \delta}$) is -0.0239 and
-0.0897, respectively. For comparison our astrometric measurement
results are listed in
Table~\ref{tabpara}.

Some debris left over from the planet formation may cause PSR~B1257+12 to be of low apparent X-ray efficiency.
It is hard to conclude whether this pulsar is low apparent X-ray efficient or not because of distance
uncertainties \citep{pkg07}. According to the X-ray measurement results from \cite{pkg07} and
our new distance result, for the 90\% confidence lower boundary of the distance 649~pc, the
X-ray efficiency of this pulsar is $9.63\times10^{-5}$. The best fitted distance 710~pc gives
an X-ray efficiency of $1.68\times10^{-4}$. So, our new VLBI result indicates that the X-ray
efficiency of PSR~B1257+12 should still be in the same range ($\sim10^{-4}-10^{-2.5}$) as
other MSPs.

As we only have 3~epochs of successful observations of PSR~J1022+1001
with the EVN, it is impossible to fit both the
distance and proper motion of this pulsar. Using the distance ($\sim$300~pc) of PSR~J1022+1001
obtained with timing method \citep{hbo04}, the two dimensional proper motions $\mu_{\rm \alpha}=-10.13$~mas/yr,
$\mu_{\rm \delta}=16.89$~mas/yr, as estimated with these 3~epochs EVN measurements.

We plan an astrometry project of more MSPs, including PSR~J1022+1001, whose model-independent distance
has not been obtained in our present work.

\acknowledgements
We are grateful to A.~Wolszczan, W.F.~Brisken, R.M.~Campbell, A.T.~Deller, B.~Zhang,
S.~Chatterjee for their kind help and suggestions. This work is partly supported by
China Ministry of Science and
Technology under State Key Development Program for Basic Research (2012CB821800),
the National Natural Science Foundation of China (grants 10625314, 11121062 and 11173046),
and the CAS/SAFEA International Partnership Program for Creative Research Teams.

\end{document}